\documentclass[a4paper,11pt]{article}
\usepackage[utf8]{inputenc}
\usepackage{amsmath}
\usepackage{amssymb}
\usepackage{graphicx}
\usepackage[sort]{cite}
\usepackage[T1]{fontenc}
\usepackage[a4paper,left=1.0in, right=1.0in,top=1.1in, bottom=1.2in]{geometry}
\usepackage[affil-it]{authblk}
\usepackage{color}
\newcommand{\be}{\begin{equation}}
\newcommand{\ee}{\end{equation}}

\newcommand{\pt}{p_{\rm T}}

\begin{document}
\title{%
{\normalsize\hspace{\fill}IFJPAN-IV-2017-29}\\[2ex]
\Large \bf Single inclusive jet production and the nuclear modification ratio\\ at very forward rapidity in proton-lead collisions\\ with $\sqrt{s_{NN}}$ = 5.02 TeV
}

\author{Marcin~Bury${}^a$, Hans Van Haevermaet${}^{b}$, Andreas Van Hameren${}^a$, Pierre Van Mechelen${}^b$,  Krzysztof~Kutak${}^a$, Mirko~Serino${}^{a,c}$
\bigskip \\
${}^a$Institute of Nuclear Physics, \\
Polish Academy of Sciences, \\ ul.\ Radzikowskiego 152, 31-342, Cracow,
Poland\\
\vspace{1cm}
${}^b$Particle Physics group, \\ University of Antwerp, \\ 
Groenenborgerlaan 171, Belgium \\
\vspace{1cm}
${}^c$ Department of Physics, \\  Ben Gurion University of the
Negev, \\ Beer Sheva 8410501, Israel
}

\maketitle

\begin{abstract}
We present calculations of single inclusive jet transverse momentum and energy spectra at forward rapidity ($5.2\!<\!y\!<\!6.6$) in proton-lead collisions with $\sqrt{s_{NN}}$ = 5.02 TeV.  
The predictions are obtained with the KaTie Monte Carlo event generator, which allows to calculate interactions within the High Energy Factorisation framework. 
The tree-level matrix element results are subsequently interfaced with the \textsc{CASCADE} Monte Carlo event generator to account for hadronisation. The effects of the saturation of the gluon density, leading to suppression of the cross section, are investigated.  
\end{abstract}

\section{Introduction}

The measurement of single inclusive forward jet production allows to study various aspects of hadron-hadron scattering.  
Jets resulting from a hard parton interaction are produced in the forward direction
if the incoming partons have a large imbalance in their longitudinal momentum component inherited from the hadrons. 
Therefore, such processes are ideal to test theoretical approaches aimed at studying both high-$x$ and low-$x$ phenomena~\cite{Gribov:1984tu,N.Cartiglia:2015gve}. 

In particular, the saturation of gluons in hadrons is still one of the challenging problems in QCD and is a subject of intensive studies (see ~\cite{Armesto:2014sma,vanHameren:2016ftb,Mantysaari:2017slo,Kotko:2017oxg,Marquet:2017bga,Ducloue:2017dit} and references therein). 
It is required by the theory in order to guarantee unitarity of the QCD evolution equations, and follows from constraints on the rate of growth of the cross section 
as the energy of the collision increases.  
Microscopically, saturation is an outcome of the competition between gluon splitting and gluon recombination processes, and can be theoretically modelled 
by nonlinear QCD evolution equations~\cite{Balitsky:1995ub,Kovchegov:1999yj,Kovchegov:1999ua,JalilianMarian:1997gr,Iancu:2001ad}. 
For nuclei, the saturation scale is expected to grow with the third inverse power of the atomic number, which is why collisions between protons and lead nuclei are a more promising environment
to look for saturation than proton-proton collisions. Phenomenological studies of various processes suggest that saturation indeed occurs in 
nature~\cite{GolecBiernat:1998js,Albacete:2010pg,Dumitru:2010iy,Adare:2011sc,Braidot:2010zh,Dusling:2013qoz,Kutak:2012rf}. 

In this paper, we present calculations of the differential cross section for the production of single inclusive jets as a function 
of transverse momentum and energy within the rapidity interval $5.2 < y < 6.6$ in proton-lead collisions at $\sqrt{s_{NN}}$ = 5.02 TeV.  
The chosen rapidity range corresponds to the acceptance of the CASTOR calorimeter, installed only on one side of the nominal interaction 
point ($-6.6 < \eta < -5.2$) of the CMS experiment~\cite{Chatrchyan:2008aa}, 
which has collected proton-proton and proton-lead collision data at the LHC at various centre-of-mass energies. 
In our setup, the phase space is defined such that in proton-lead collisions the proton hosting the high-$x$ initial state parton 
moves towards the negative rapidity hemisphere, where the CASTOR detector resides. 
This allows to probe the structure of the lead ion, that moves away from CASTOR and hosts a very low-$x$ parton.
All results are presented in the centre-of-mass frame and need to be boosted to the laboratory frame for comparisons with data measurements.

The spectra are calculated using High Energy Factorisation (HEF), extending this framework to allow for nonlinearities in one of the colliding hadrons~\cite{Catani:1990eg,Deak:2009xt}. 
In this approach, matrix elements for single inclusive jet production can be given at leading order as a $2 \to 1$ process, with one of the incoming partons being off-shell. 
This is in contrast to collinear factorisation, where the matrix element for the $2 \to 1$ process vanishes at leading order
and, thus, one has to use matrix elements of higher order in $\alpha_{\rm S}$ to account for the finite transverse momentum of the jet. 

%%%%%%%%%%%%%%%%%%%%%%%%%%%%%%%%%%%%%%%%
\section{Single inclusive jet production in High Energy Factorisation}
%%%%%%%%%%%%%%%%%%%%%%%%%%%%%%%%%%%%%%%%

The single inclusive jet production process can be schematically written as
\begin{equation}
{\rm A} + {\rm B} \rightarrow {\rm a} + {\rm b} \rightarrow \text{jet} + {\rm X}
\end{equation}
where A and B are the colliding hadrons, each of which provides a parton, 
respectively $a$ and $b$, and $X$ corresponds to undetected real radiation. 
The beam remnants from the hadrons A and B are understood to be implicitly included in the above equation.

The longitudinal kinematic variables can be expressed as
\begin{equation}
x_1 = \frac{1}{\sqrt{s_{NN}}}\, p_{\rm T}\, e^{y}, \qquad x_2 = \frac{1}{\sqrt{s_{NN}}}\, p_{\rm T}\, e^{-y},
\end{equation}
with $s_{NN}=(p_{\rm A}+p_{\rm B})^2$  the total squared energy of the colliding nucleons, while $y$ and $p_{\rm T}$ are the rapidity and transverse momentum of the leading final state jet, respectively.

The HEF\footnote{The formula is at leading order accuracy. There is ongoing activity to advance it to NLO 
level~\cite{Chirilli:2011km,Altinoluk:2014eka,Iancu:2016vyg}. For a review see~\cite{Stasto:2016wrf}, and for recent applications of HEF framework to other processes 
see~\cite{Deak:2010gk,Deak:2011ga,Deak:2009ae,Deak:2011gj,Kutak:2016mik,Kutak:2016ukc,Luszczak:2016csq,Baranov:2015yea,Dooling:2014kia,Sapeta:2015gee,Kotko:2016lej,Cisek:2017ikn} 
and references therein.} formula applicable for the rapidity range that we address in this paper reads~\cite{Dumitru:2005gt}
\begin{equation}
\begin{split}
\frac{{\rm d}\sigma}{{\rm d}y\, {\rm d}p_{\rm T}} = 
\frac{1}{2}\frac{\pi\, p_{\rm T}}{(x_1 x_2 s_{NN})^2} & 
\bigg[
\sum_{\rm q(\bar q)}\overline{|{\cal M}_{\rm g^*q(\bar q)\to q (\bar q)}|}^2 \, x_1\, f_{\rm q(\bar q)/A}(x_1,\mu^2)\, {\cal F}^F_{\rm g^*/B}(x_2,p_{\rm T}^2,\mu^2)  \\ & 
\hspace{5mm}
+ \overline{|{\cal M}_{\rm g^*g\to g}|}^2 \, x_1\, g_{\rm g/A}(x_1,\mu^2)\, {\cal F}^A_{\rm g^*/B}(x_2,p_{\rm T}^2,\mu^2)) 
\bigg] \,,
\label{eq:int-phi}
\end{split}
\end{equation}
where ${\cal F}^{F(A)}$ is the unintegrated gluon density parametrising the target, 
which can be a proton or a lead nucleus, respectively in the fundamental and adjoint color representations. 
These functions depend on the longitudinal momentum fraction $x$, the transverse momentum $\pt$ and on some hard scale $\mu$.
The matrix elements squared respectively represent the scattering of an off-shell gluon with an on-shell quark
and an off-shell gluon with an on-shell gluon, and are averaged over initial state helicities 
(indicated by the bar) and summed over final state helicities~\cite{vanHameren:2012if, Bury:2016cue}.  
As input in the formula above we use the following unintegrated gluon densities:
\begin{itemize}
\item KS-linear, which is a solution of the momentum space version of the extended BFKL equation~\cite{Kuraev:1977fs,Balitsky:1978ic,Kuraev:1976ge} 
and, as such, does not include saturation effects. It is obtained following the Kwiecinski-Martin-Stasto (KMS) prescription~\cite{Kwiecinski:1997ee}.
The extended splitting function respectively reduces to the BFKL splitting function in the low $z$ limit and to the DGLAP splitting function for strong ordering in transverse momentum.
The prescription uses a specific kinematical constraint that limits the phase space for real gluon emission. 
This assures that virtuality of the exchanged gluon is dominated by its transverse momentum. It also takes into account the contribution of splitting to quarks and effects from the running coupling strength.
The gluon density is normalised to the number of nucleons.
\item KS-nonlinear, which is a solution of the momentum space version of the BK equation~\cite{Balitsky:1995ub,Kovchegov:1999yj} 
with modifications according to the KMS prescription \cite{Kutak:2003bd}. As such, it accounts for saturation effects.
The initial conditions of the equation are modified to account for the difference between a proton and a nucleus as explained in~\cite{Kutak:2012rf}.  
This gluon density is used for both color representations, to stay in HEF approximation.
Similarly as above the gluon density is normalised to the number of nucleons.
\item KMRW-lead, obtained using the Kimber-Martin-Ryskin-Watt (KMRW) prescription. 
This is a gluon density obtained from a collinear gluon density by resumming virtual and soft emissions 
via the Sudakov form factor~\cite{Kimber:1999xc,Watt:2003mx}. 
The underlying parton density functions (PDF) from which we construct the transverse momentum dependent (TMD) PDF are the nPDFs obtained in \cite{Kovarik:2015cma,Eskola:2016oht}.
\end{itemize}
As for the high-$x$ parton, which is taken to be on the mass-shell, we use the CT10 collinear PDFs~\cite{Guzzi:2011sv}.
As already mentioned, in our calculations we take $y_{\rm min} = 5.2$ and $y_{\rm max} = 6.6$, which corresponds to the acceptance of the CASTOR calorimeter. 

%%%%%%%%%%%%%%%%%%%%%%%%%%
\section{Numerical results}
%%%%%%%%%%%%%%%%%%%%%%%%%%

All samples used in this section are obtained with the KaTie Monte Carlo event generator, which can produce tree-level matrix element calculations in HEF~\cite{vanHameren:2016kkz}. 
In this particular case, $gg^* \rightarrow g$ and $qg^* \rightarrow q$ processes at $\sqrt{s_{\text{NN}}} = 5.02$ TeV 
are generated with the initial gluon being off-shell. The minimum $\pt$ of the final-state jet is set to $1.4$ GeV. 
The renormalisation and factorisation scale is set to be the transverse momentum of the final state jet. Finally, the aforementioned CT10 NLO is 
used as collinear PDF for the on-shell parton, and the specific KS-linear, KS-nonlinear and KMRW-lead unintegrated gluon PDFs for the off-shell parton.
The resulting samples are analysed with Rivet~\cite{Buckley:2010ar}
where the jets are clustered with FastJet~\cite{Cacciari:2011ma} using the anti-$k_{\rm T}$ algorithm with distance parameter equal to 0.5~\cite{Cacciari:2008gp}.
To obtain the hadron level samples, the output from KaTie is fed to the CASCADE 2.4.13 Monte Carlo event generator~\cite{Jung:2010si}
that employs an adopted version of the Lund string fragmentation model (as used in Pythia) to account for hadronisation.

%%%%%%%%%%%%%%%%%%%%%%%%%%%%
\subsection{Transverse momentum and energy spectra}
%%%%%%%%%%%%%%%%%%%%%%%%%%%%

\begin{figure}[t!]
\centering
\includegraphics[width=0.49\textwidth]{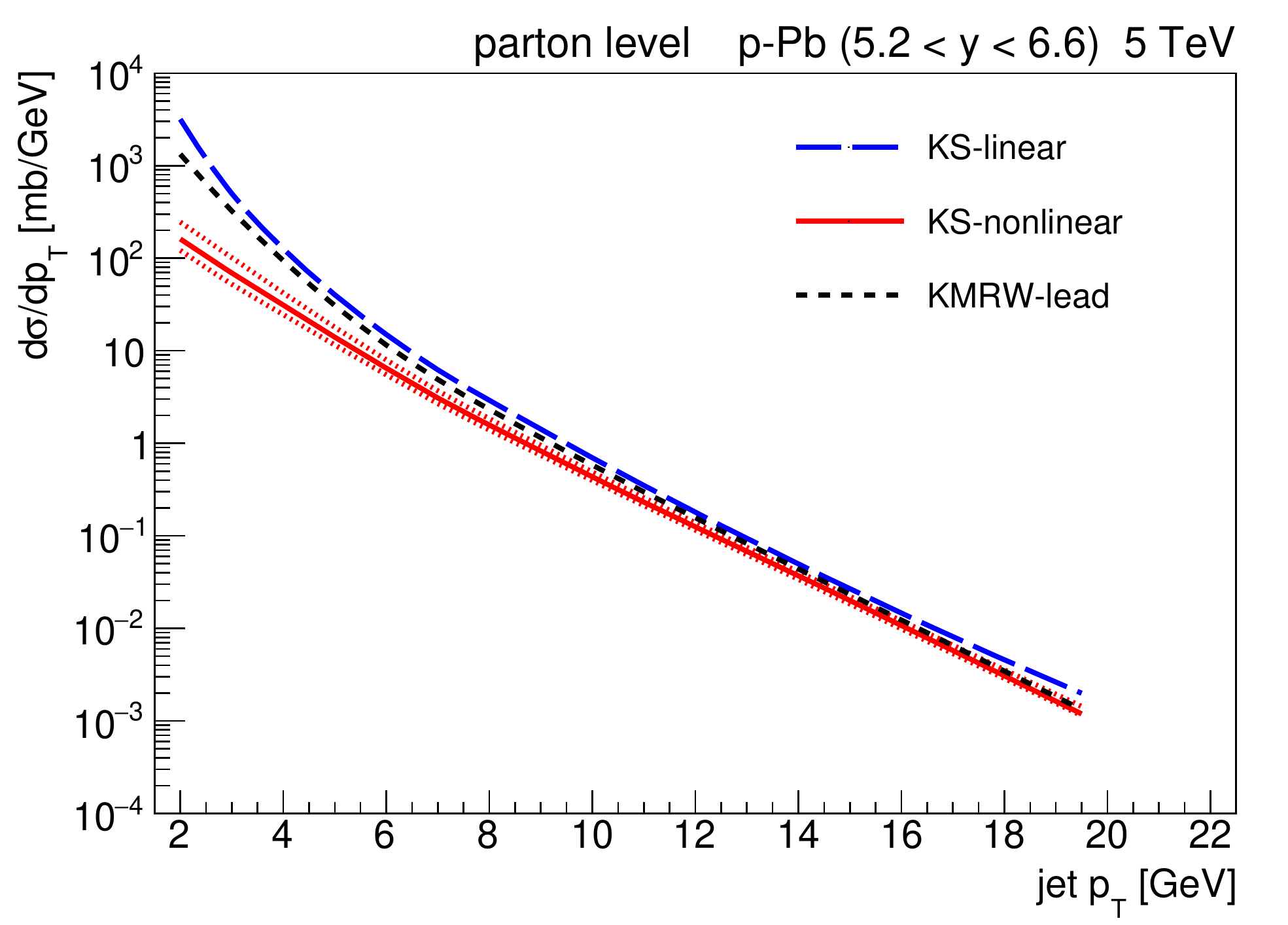}
\includegraphics[width=0.49\textwidth]{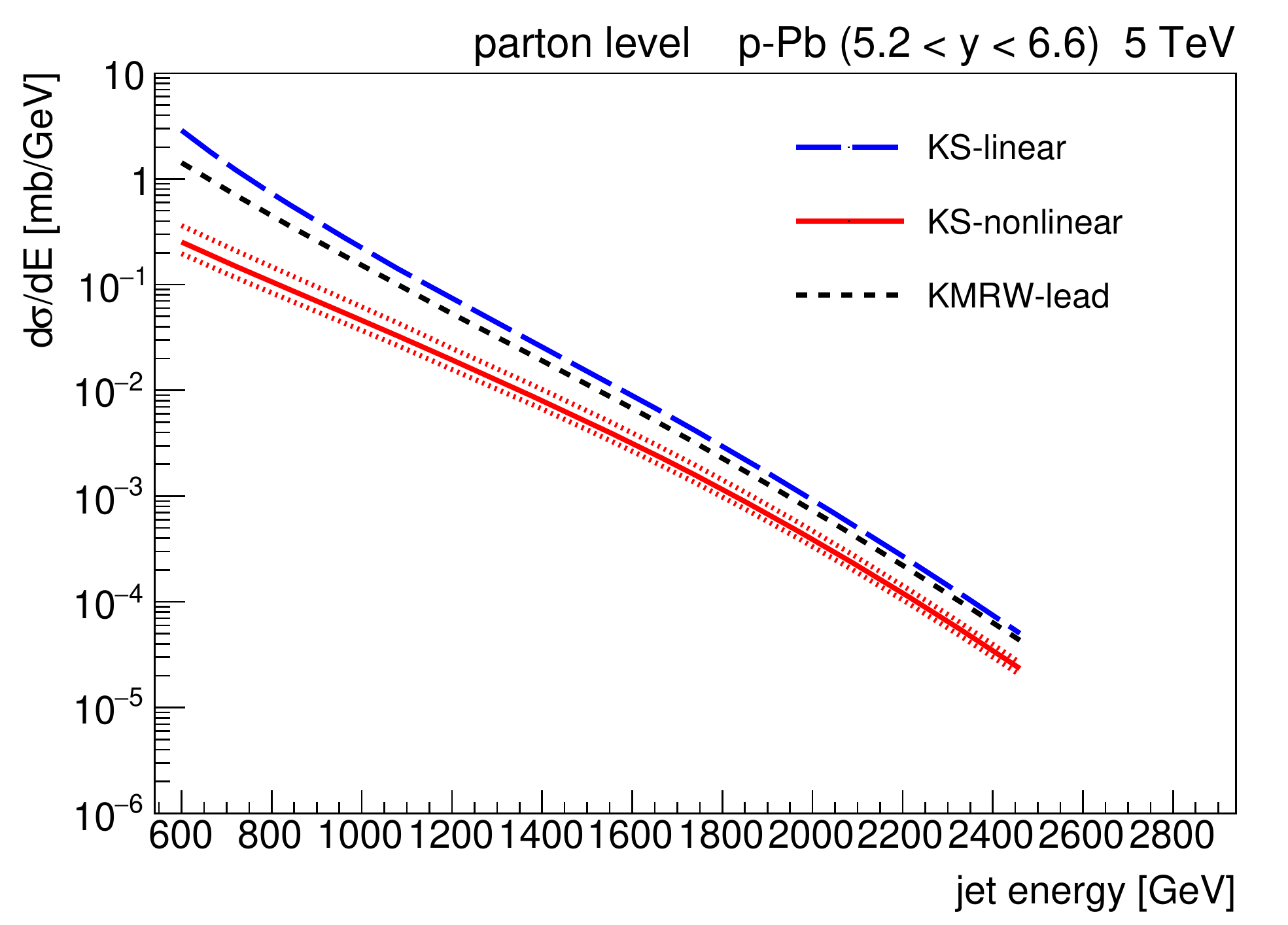}
\caption{Parton level predictions of KaTie for KS-linear, KS-nonlinear and KMRW-lead gluon densities. Differential jet cross sections as a function of jet $\pt$ (left) and energy (right) are presented for proton-lead interactions. The fine dotted lines represent the upper and lower uncertainty in the nonlinear suppression factor.}
\label{fig:spectra_pA}
\end{figure}

\begin{figure}[h!]
\centering
\includegraphics[width=0.49\textwidth]{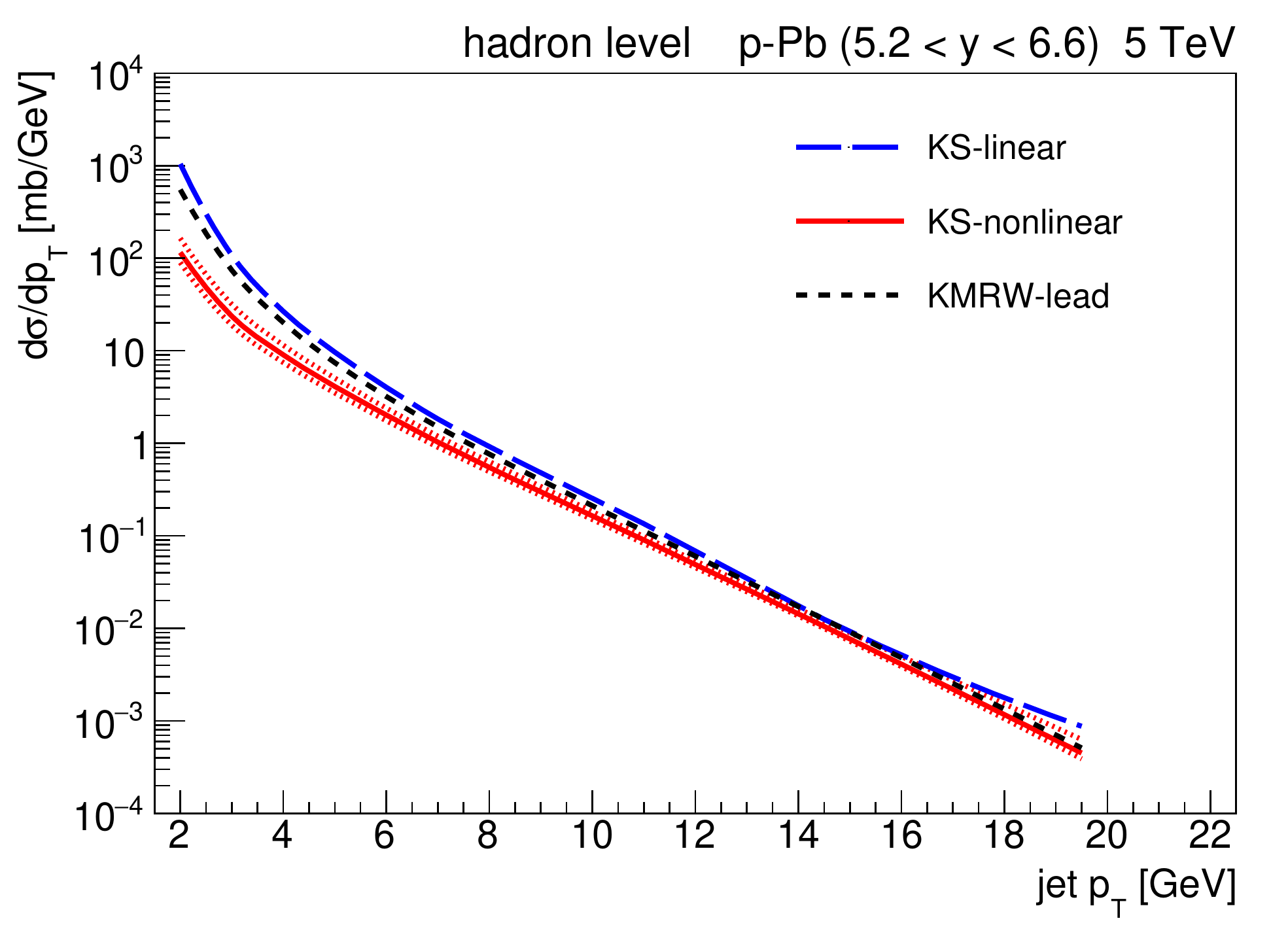}
\includegraphics[width=0.49\textwidth]{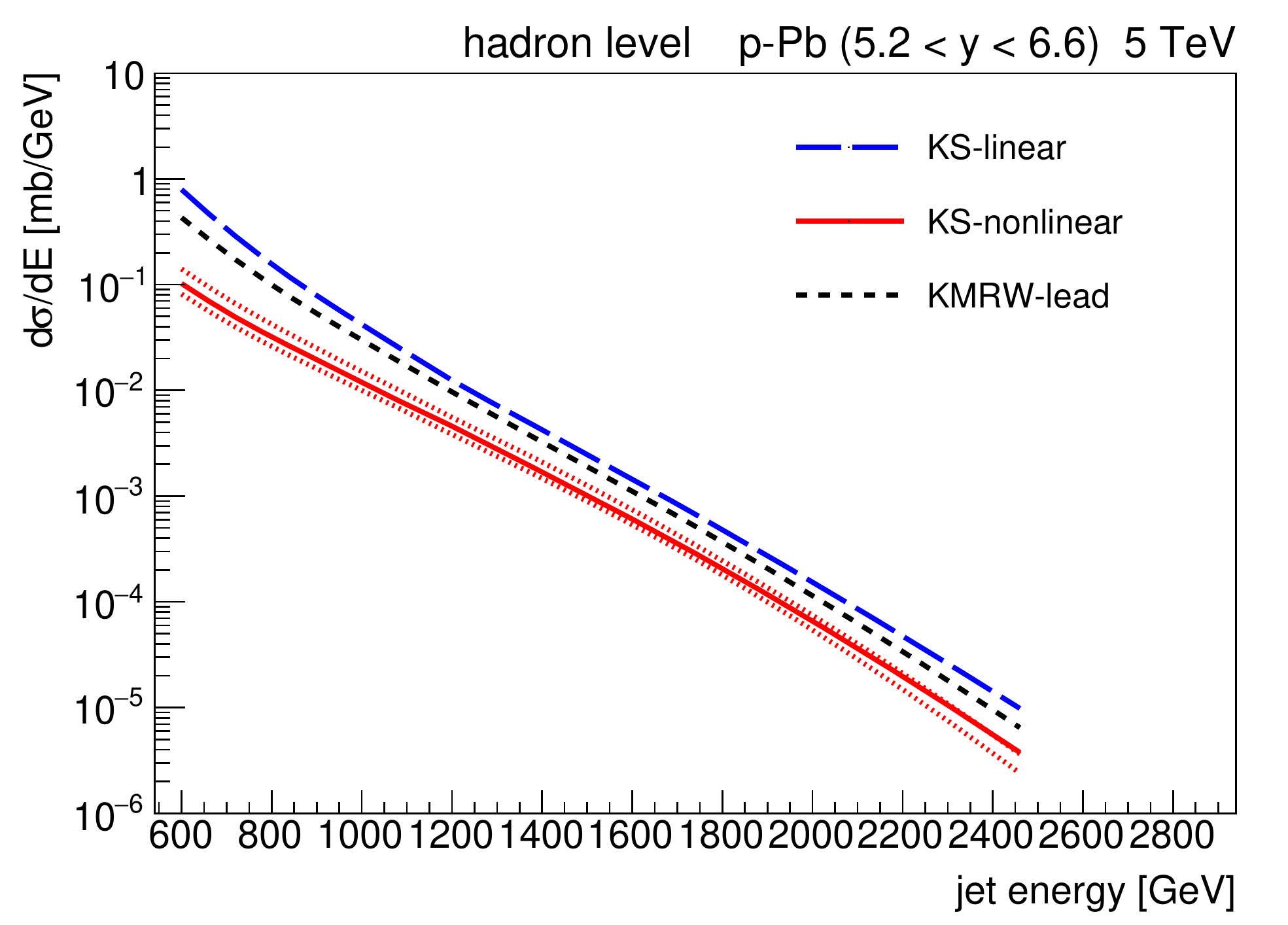}
\caption{Hadron level predictions of KaTie+CASCADE for KS-linear, KS-nonlinear and KMRW-lead gluon densities. Differential jet cross sections as a function of jet $\pt$ (left) and energy (right) are presented for proton-lead interactions. The fine dotted lines represent the upper and lower uncertainty in the nonlinear suppression factor.}
\label{fig:spectra_pA_had}
\end{figure}

The observable that conveniently reveals the partonic dynamics in which we are interested is the inclusive jet transverse momentum spectrum: 
in particular, when the jet is produced in the forward direction, its $\pt$ has to be sufficiently small for it to potentially carry information about the saturation phenomenon. 
In Fig.\@~\ref{fig:spectra_pA} (left) we plot the parton level cross section as a function of $\pt$, while in Fig.\@~\ref{fig:spectra_pA} (right) it is presented as a function of the energy of the jet. 
Predictions obtained using KaTie with KS-linear, KS-nonlinear and KMRW-lead parton densities are compared. 
The result with the KS-nonlinear PDF is obtained when the nonlinear suppression term is multiplied by a parameter equal to 0.75. 
Since this is a free parameter that needs input from data, additional scenarios 
where the nonlinear suppression term is multiplied by 0.5 and 1.0 are included as fine dotted curves 
in Fig.\@~\ref{fig:spectra_pA} (and all other following figures in the paper).
Although this choice is arbitrary, it does indicate the sensitivity of the observable if the nonlinear suppression is altered by $\pm 33\%$.
We see that, as expected, at lower $\pt$ and lower energy the saturation effects can be substantial, since the KS-nonlinear result is suppressed as compared to KS-linear. 
In addition we see that KMRW-lead, which follows from DGLAP evolution equations and does not account for nonlinear effects during the evolution but accounts for nuclear shadowing \cite{Armesto:2006ph}, is very close to KS-linear.
In Fig.\@~\ref{fig:spectra_pA_had} we add hadronisation effects by combining the output of KaTie with CASCADE (following methods developed in~\cite{Bury:2017jxo}), 
which leads to a significant decrease of the overall cross section and slightly changes the shape of the distributions as well. 
Nevertheless, a clear difference in the predictions remains at lower $\pt$ and energy values, 
which makes the observable suitable for comparisons with measurements from data corrected to particle level.

%%%%%%%%%%%%%%%%%%%%%%%%%%%%%%%
%%%%%%%%%%%%%%%%%%%%%%%%%%%

%%%%%%%%%%%%%%%%%%%%
\subsection{Nuclear modification ratio}
%%%%%%%%%%%%%%%%%%%%

In order to quantify the strength of nonlinearities as one goes from proton-proton to proton-lead, 
it is convenient to calculate a quantity called the nuclear modification ratio $R_{pA}$.
For a generic observable ${\cal O}$, it is defined  as
\begin{equation}
R_{pA}=\frac{\sigma_{pPb}({\cal O})}{A\, \sigma_{pp}({\cal O})}.
\end{equation}
On one hand, in absence of any effects like saturation in the low-$x$ evolution equations, the ratio would be just consistent with unity (since one considers gluon dominated observables). 
On the other hand, if saturation effects are present, they would be visible to us in the nuclear modification ratio as a deviation from unity within some range, 
for instance in the transverse momentum spectra of the measured jets. 
In the low-$x$ approach the suppression is directly linked to a denser partonic system (and therefore larger contribution of the nonlinear term) as one goes from proton to lead.
In the DGLAP approach the possible suppression is due to shadowing, which is implemented by fitting parton densities to data without accounting for any additional dynamical effects that may happen when going from a proton to a nucleus \footnote{In general, in absence of any nuclear effects and saturation, the small deviation from unity is due to the difference between proton and neutron PDFs that contribute to a nuclear PDF.}.  
 
\begin{figure}[t!]
\centering
\includegraphics[width=0.49\textwidth]{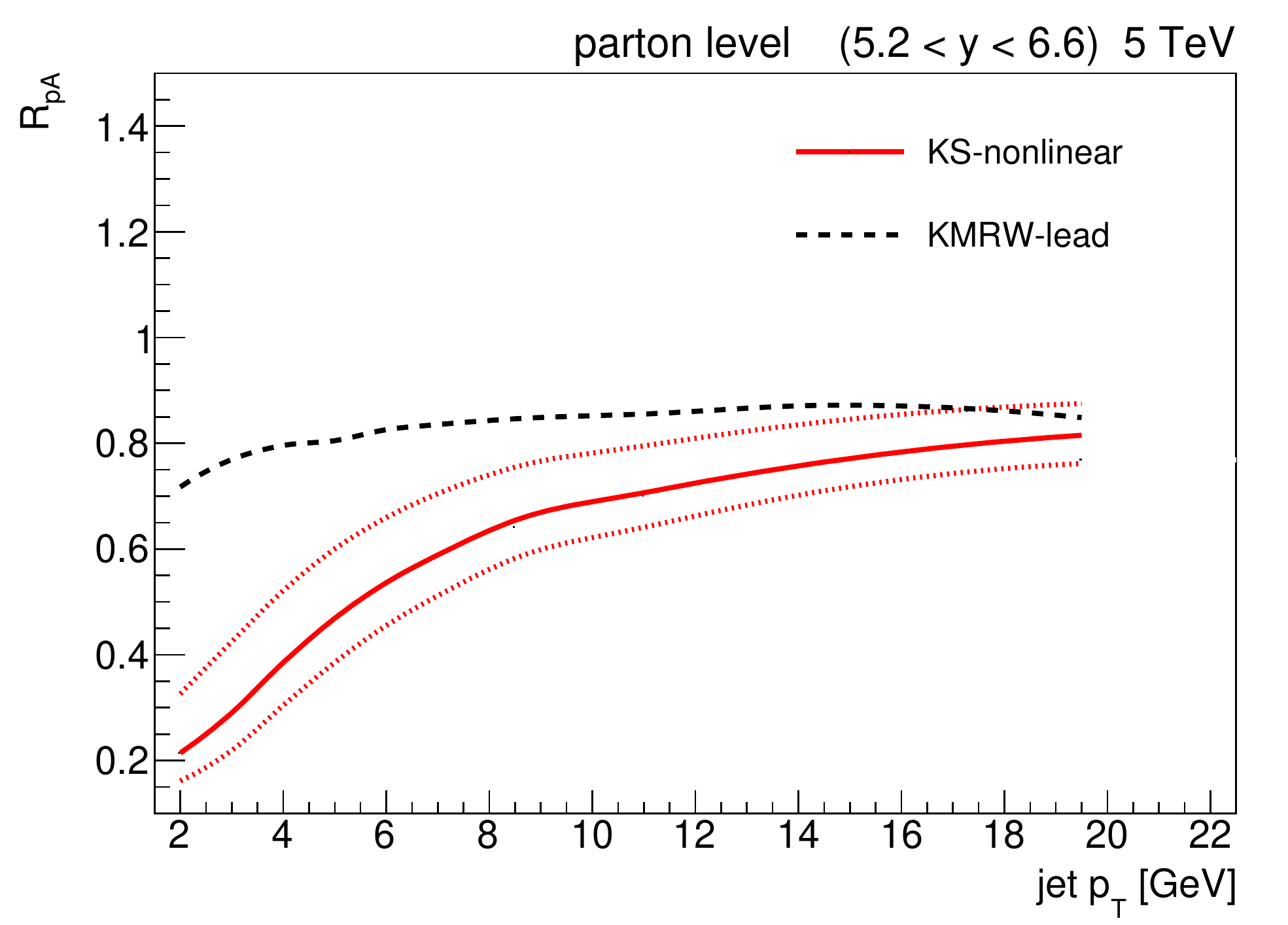}
\includegraphics[width=0.49\textwidth]{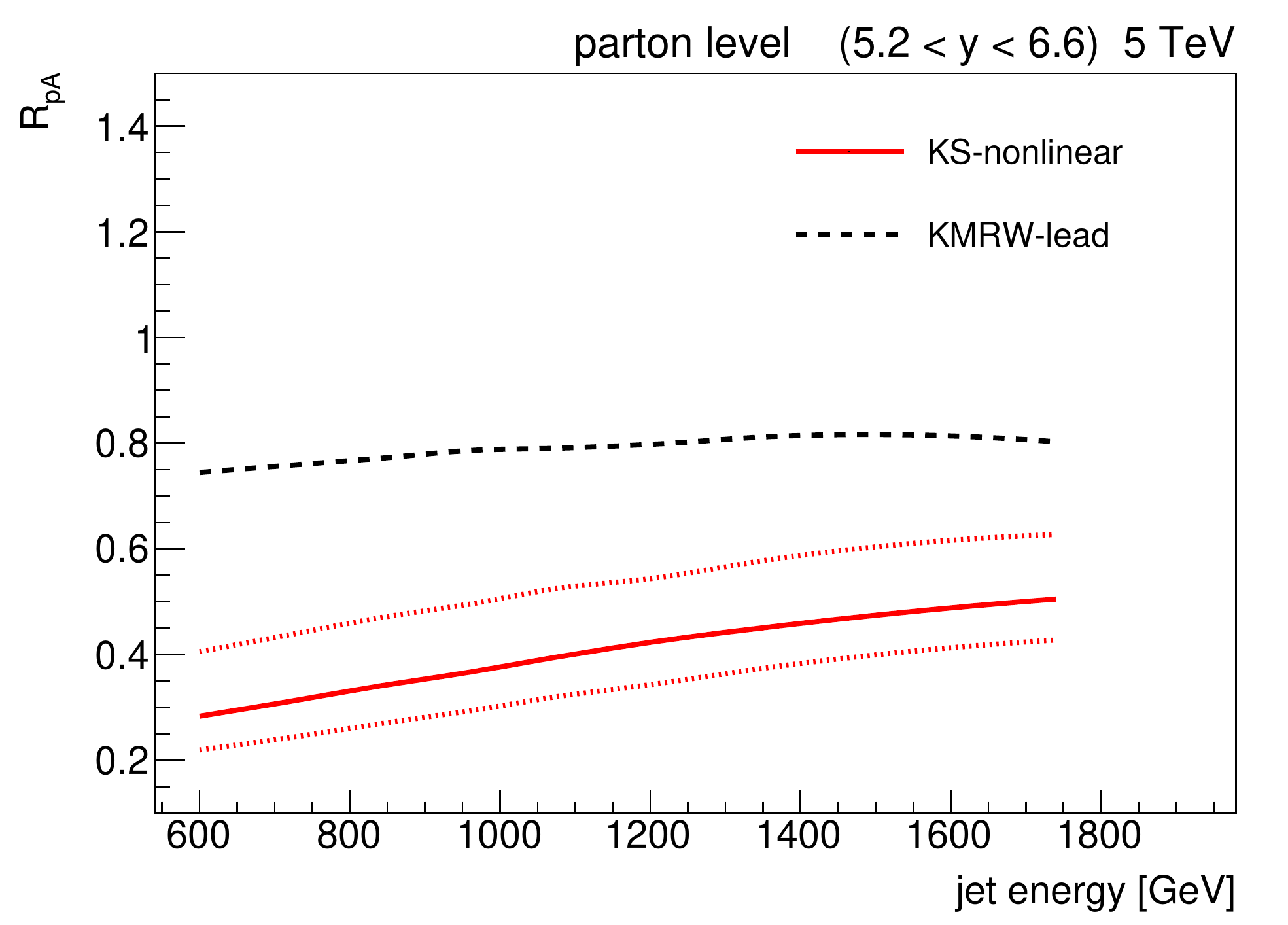}
\caption{Parton level predictions of the nuclear modification ratio $R_{pA}$ as function of the jet $p_T$ (left) and jet energy (right). The fine dotted lines represent the upper and lower uncertainty in the nonlinear suppression factor.}
\label{fig:spectra_RpA_parton}
\end{figure}

\begin{figure}[t!]
\centering
\includegraphics[width=0.49\textwidth]{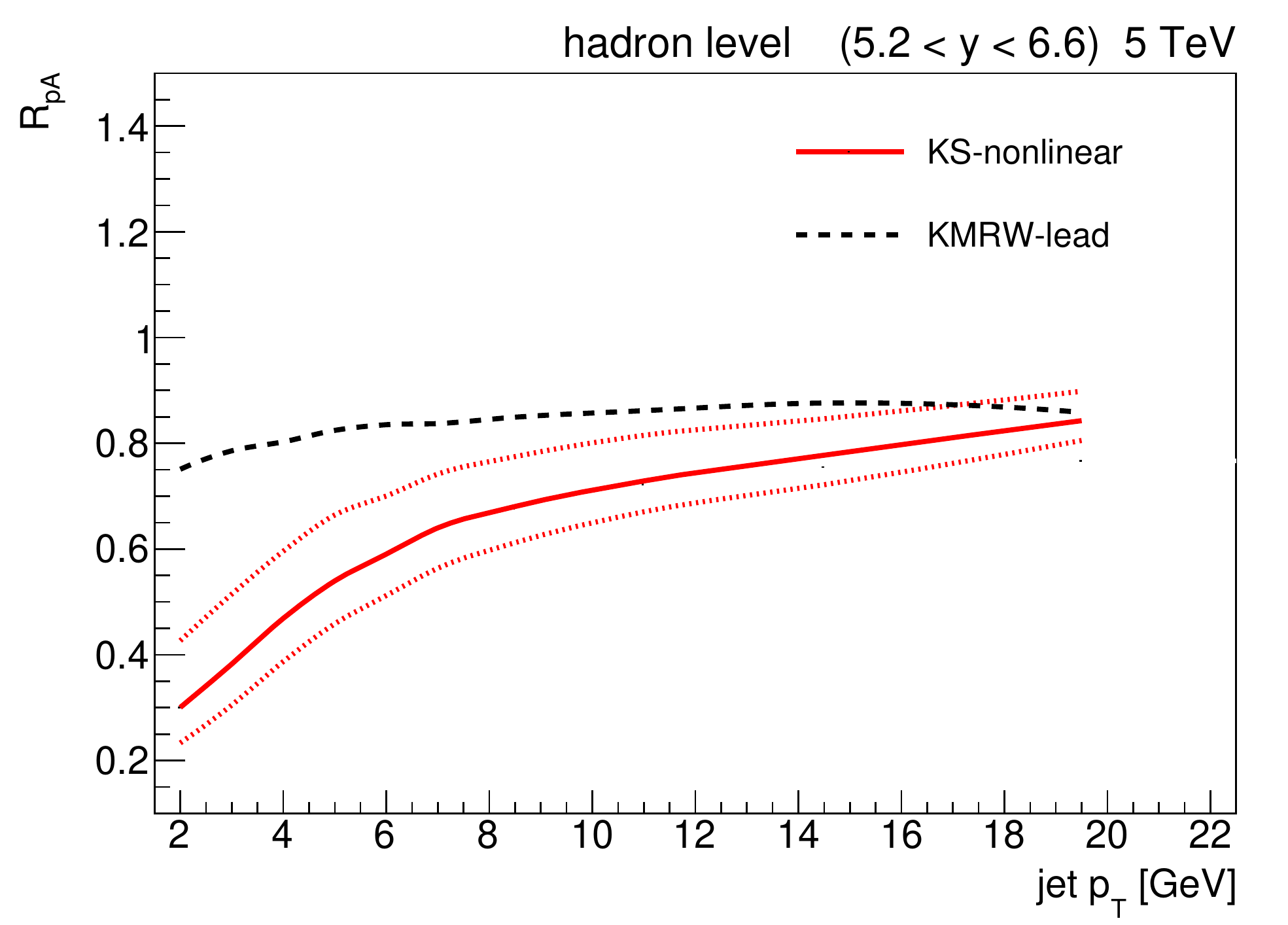}
\includegraphics[width=0.49\textwidth]{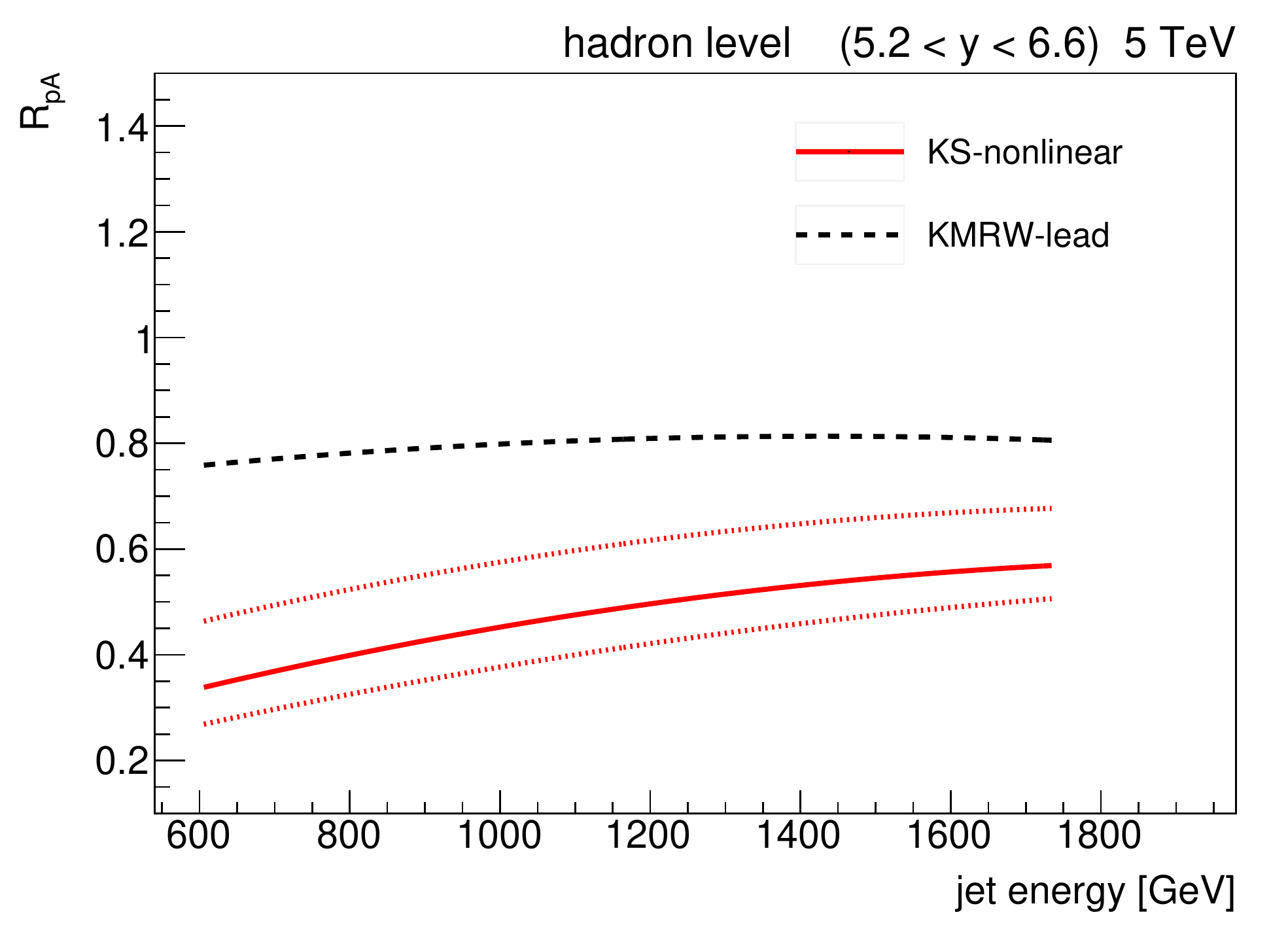}
\caption{Hadron level predictions of the nuclear modification ratio $R_{pA}$ as function of the jet $p_T$ (left) and jet energy (right). The fine dotted lines represent the upper and lower uncertainty in the nonlinear suppression factor. }
\label{fig:spectra_RpA_hadron}
\end{figure}

In Figs.\@~\ref{fig:spectra_RpA_parton} and \ref{fig:spectra_RpA_hadron} we compare parton and hadron level predictions 
of the nuclear modification ratios as obtained using KS-nonlinear and KMRW-lead parton densities. 
It shows a significant suppression for KS-nonlinear at low values of $\pt$, which indicates that the saturation of the gluon density in lead is large compared to the saturation in the proton.
At high values of $\pt$ the results obtained with KS-nonlinear and KMRW-lead converge, which shows that nonlinear suppression is negligible in this region. 
Both do not converge to unity however, which can imply that other suppression effects coming from the nuclei are present. 
The results obtained using the KMRW-lead gluon density exhibits a different, more constant, behaviour as it does not include saturation effects. However, the nuclear shadowing can, even at  $\pt=10$ GeV, be substantial since at the considered energies the nuclear PDF is probed at $x=10^{-5}$. 

The nuclear modification ratio as a function of jet energy shows a similar behaviour, 
although the effects are somewhat washed out, leading to an overall different slope and normalisation of the KS-nonlinear and KMRW-lead predictions.
Looking at the fine dotted curves in Figs.\@~\ref{fig:spectra_RpA_parton} and \ref{fig:spectra_RpA_hadron}, it is seen that the uncertainty due to saturation is large, 
indicating a high sensitivity of these observables to saturation effects. 
Even though the variations can be significant, there is always a clear difference with respect to KMRW-lead, which does not include such effects. 
Therefore a measurement of the nuclear modification factor of forward, low $\pt$ jets is ideal to disentangle linear from nonlinear effects 
and to constrain the amount of suppression in the cross section.

\begin{figure}[t!]
\centering
\includegraphics[width=0.49\textwidth]{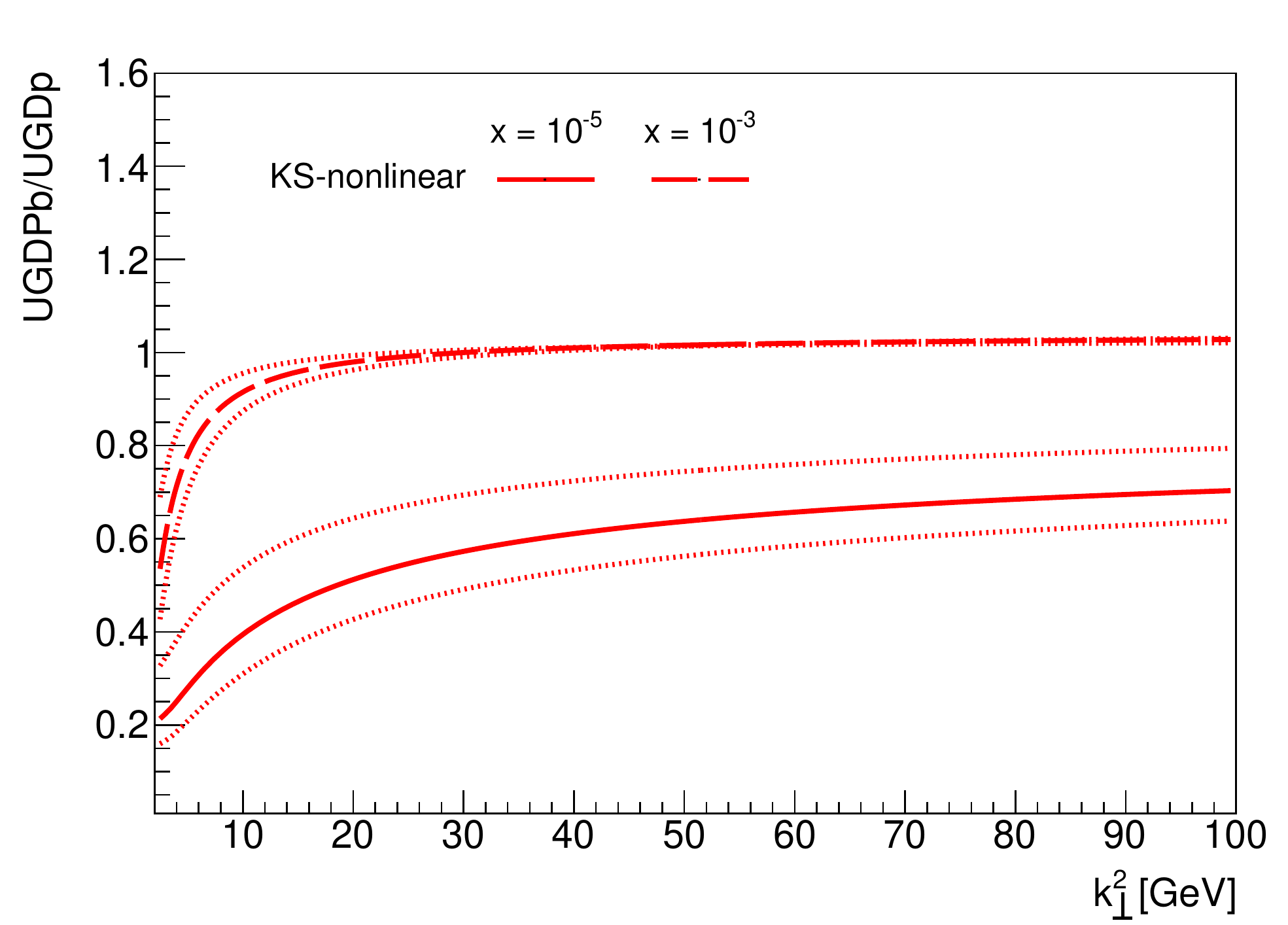}
\includegraphics[width=0.49\textwidth]{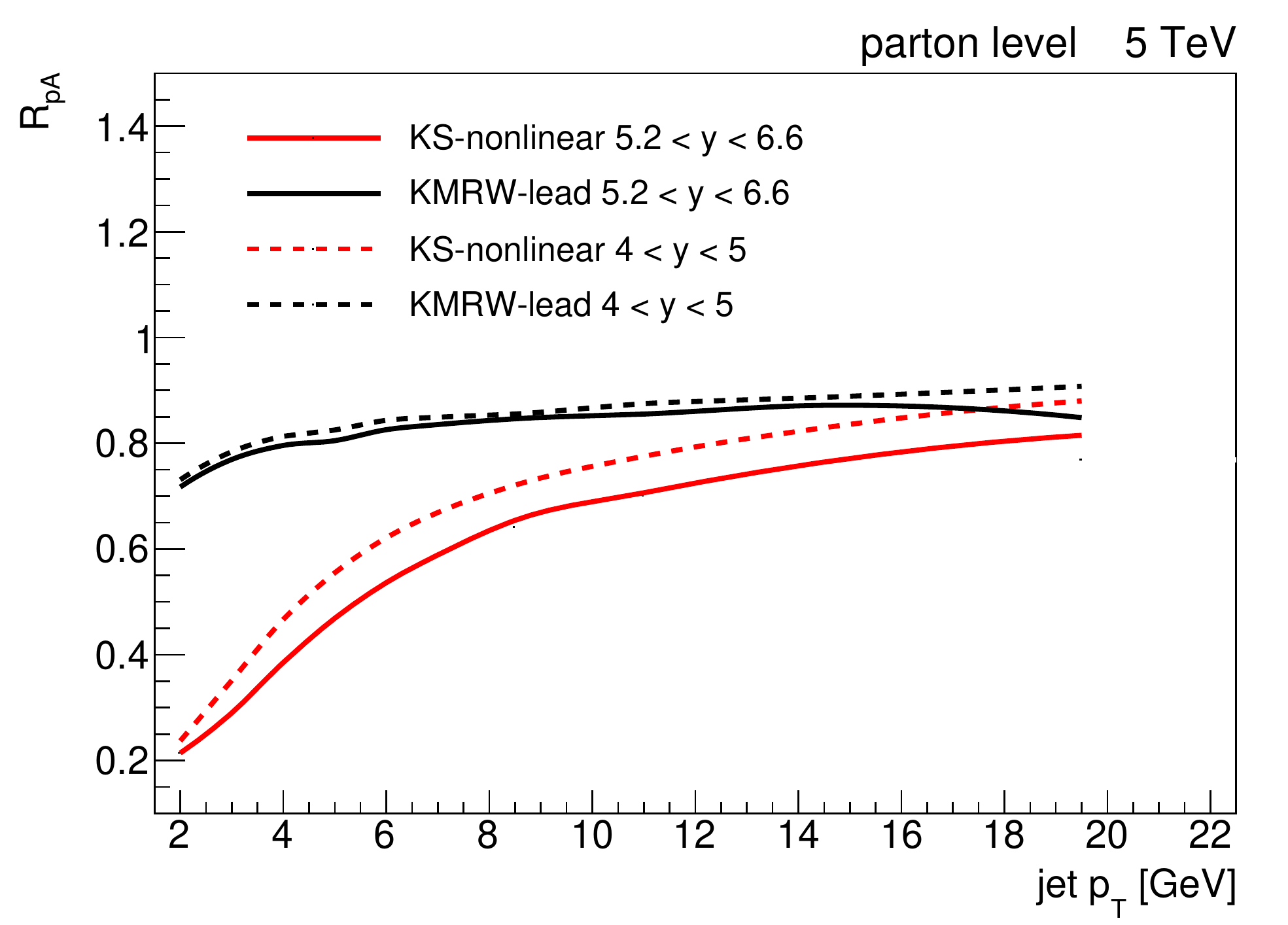}
\caption{Ratio of KS-nonlinear unintegrated gluon densities for lead and proton evaluated at different values of $x$. 
The fine dotted lines represent the upper and lower uncertainty in the nonlinear suppression factor (left). 
Parton level comparison of the nuclear modification ratio $R_{pA}$ for jets within $5.2 < y < 6.6$ and, 
more centrally, in $4.0 < y < 5.0$ for both KS-nonlinear and KMRW-lead (right).}
\label{fig:spectra_RpA_PDFs}
\end{figure}

In order to better understand this suppression mechanism of the cross section, 
we also plot the ratio of the unintegrated gluon densities for lead ions (UGDPb) and protons (UGDp) as a function of $k_\perp^{2}$ 
in Fig.\@~\ref{fig:spectra_RpA_PDFs} (left), evaluated at different values of $x$. 
We see that at larger values, $x = 10^{-3}$, the suppression is much reduced and the ratio converges quickly to unity. 
At very low values, $x = 10^{-5}$, we see a similar behaviour as reported before. 
This is consistent, since the production of low $\pt$ jets within $5.2 < y < 6.6$ reaches values of $x$ as low as $10^{-6}$. 

In addition, Fig.\@~\ref{fig:spectra_RpA_PDFs} (right) shows parton level predictions of $R_{pA}$ for jets within the default rapidity range, $5.2 < y < 6.6$,range 
and for jets that are produced more centrally in $4.0 < y < 5.0$, for both KS-nonlinear and KMRW-lead parton densities. 
This confirms the dependence of the nonlinear behaviour on different probed rapidity regions, 
and shows that parton densities that do not incorporate nonlinear dynamics such as KMRW-lead are less sensitive. 

%%%%%%%%%%%%%%%%%%%%%%%%%%%%%%%%%%
%%%%%%%%%%%%%%%%%%%%%%%%%%%%%%%%%%
%%%%%%%%%%%%%%%%%%%%%%%%%%%%%%%%%%

%%%%%%%%%%%%
\section{Conclusions}
%%%%%%%%%%%%

In this paper we have calculated transverse momentum and energy spectra of single inclusive forward jets in the rapidity region of $5.2 < y < 6.6$, which corresponds to the CASTOR detector acceptance at the CMS experiment.
The HEF calculations have been performed using the KaTie Monte Carlo event generator supplemented with KS-linear, KS-nonlinear, and KMRW-lead parton densities,
and interfaced to the CASCADE Monte Carlo event generator in order to account for hadronisation effects. 
We observe that the energy and transverse momentum spectra of KS-linear are overall consistent with KMRW-lead spectra. 
The nonlinear dynamics as encoded in KS-nonlinear distributions predicts a suppression of the cross section for values of $\pt$ smaller than 8 GeV. 
The energy spectrum, which can be measured in the CASTOR calorimeter, is affected by the nonlinear effects in the whole range.
We also calculated nuclear modification ratios that measure the change of the dynamics as one increases the nuclear mass number. 
A clear difference between linear and nonlinear evolutions is observed with decreasing $\pt$ or $x$, as expected from saturation effects. 
In order to improve on this, one needs formal calculations with higher order accuracy both of the hard matrix elements~\cite{Iancu:2016vyg} 
as well as to account for first principles calculations of resummed higher order corrections to the gluon density including effects like: 
collinear resummation \cite{Iancu:2015vea}, running coupling and quarks contribution relevant at moderate and large $x$~\cite{Baranov:2017tig,Hautmann:2017fcj}. 
This is also because the NLO corrections~\cite{Balitsky:2008zza} introduce instabilities, as pointed out in~\cite{Lappi:2015fma}. 
Furthermore, a gluon density valid in the whole kinematical regime would increase the predictivity of the theory. 
Progress in the latter direction can be achieved once a program along the lines of~\cite{Hentschinski:2017ayz} is completed.

%%%%%%%%%%%%%%%%%%%%%
\section*{Acknowledgements}
%%%%%%%%%%%%%%%%%%%%%

This research has been supported by a common FWO-PAS VS.070.16N research grant.
MS is also partially supported by the  Israeli Science Foundation through grant 1635/16, 
by the BSF grants 2012124 and 2014707, by the COST Action CA15213 THOR
and by a Kreitman fellowship from the Ben Gurion University of the Negev. HvH is a Postdoctoral Fellow of the Research Foundation - Flanders (FWO).
AvH was partially supported by a grant of the National Science Center, Poland, No.~2015/17/B/ST2/01838.
The authors would like to thank Francesco Hautmann, Hannes Jung, Piotr Kotko and Aleksander Kusina for useful discussions.

\end{document}